\documentclass{epl}

\title{Mobile particles in an immobile environment: Molecular
Dynamics simulation of a binary Yukawa mixture}
\shorttitle{Mobile particles in an immobile environment}
\author{N. Kikuchi \and J. Horbach}
\institute{
Institut f\"ur Physik, Johannes Gutenberg--Universit\"at Mainz, Staudingerweg 7,
D--55099 Mainz, Germany}


\begin{document}

\maketitle

\begin{abstract}
Molecular dynamics computer simulations are used to investigate the
dynamics of a binary mixture of charged (Yukawa) particles with a
size--ratio of 1:5. We find that the system undergoes a phase transition
where the large particles crystallize while the small particles remain
in a fluid--like (delocalized) phase. Upon decreasing temperature below
the transition, the small particles become increasingly localized
on intermediate time scales. This is reflected in the incoherent
intermediate scattering functions by the appearance of a plateau with
a growing height. At long times, the small particles show a diffusive
hopping motion. We find that these transport properties are related
to structural correlations and the single--particle potential energy
distribution of the small particles.
\end{abstract}
\section{Introduction}
The dynamics of fluid particles in an immobile environment has
been of fundamental interest to understand anomalous transport
processes in confined geometry and porous media~\cite{confine}.
Examples include different types of Lorentz gas models
\cite{klages00,moreno04a,moreno04b,hoefling06}, polymers in
quenched disorder \cite{stepanow92,cugliandolo96,milchev04},
hard--sphere mixtures \cite{krako05} and ion--conducting silicates
\cite{meyer04,voigtmann06}. Simple model systems that provide a
time--scale separation of transport properties among different species
are mixtures of small and large particles at high densities. While it
is difficult to experimentally realize such systems on an atomistic
scale (for an exception see Ref.~\cite{egelstaff90}), it is possible
to carry out experiments on colloidal suspensions that contain
disparately--sized particles. About a decade ago, Imhof and Dhont
\cite{imhof95a,imhof95b,imhof97} performed dynamic light scattering
experiments on a binary mixture of colloidal silica particles with a
size ratio of 1:9.3. The effective interactions between colloids are
hard--sphere--like, hence phase behavior and transport properties are
governed by packing effects.  An interesting finding of Imhof and Dhont
is the existence of different phases where the large particles exhibit a
structural arrest, yet the small particles are still mobile. The simplest
case of such a phase could be one consisting of mobile small fluid-like
particles in a crystalline matrix of large particles. It can be realized,
at least, in an intermediate regime below the freezing transition.

In this work, we use molecular dynamics simulations to study a binary
mixture of charged particles with a size--ratio of 1:5.  Similar to
the experiments by Imhof and Dhont \cite{imhof95a,imhof95b,imhof97},
the system exhibits a phase transition from a fluid to a mixture of
crystalline large particles and fluid--like small particles. When
temperature is sufficiently lowered, the small particles are localized
and crystallize, eventually, in a sublattice with respect to the
large--particle lattice. In the temperature regime between the freezing
transition and the final crystallization into a sublattice, the movement
of the small particles can be characterized by a hopping motion in an
external periodic potential (created by the large particles). Similar to
supercooled liquids, a ``$\beta$ relaxation regime'' can be identified
in the incoherent intermediate scattering functions $F_{\rm s}^{\rm
(s)}(q,t)$ (with $q$ the wavenumber and $t$ the time) of the small
particles. These functions show a plateau region between the microscopic
regime and the long--time decay to zero. At a given wavenumber, the height
of the plateau increases with decreasing temperature. This indicates that
the particles are more and more localized.  Thus, the small particles
display a kind of delocalization--to--localization transition around
the liquid--to--crystal transition of the large particles.  Although
the effective packing fraction of the small particles, considered in
this work, is relatively small, structural correlations among the small
particles occur, associated with a complex distribution of single particle
potential energies.

The system considered in this work shows various nontrivial features
that might be generic for the dynamics in confined geometry and
porous media, especially when Coulomb--like interactions become
important. It can be also realized experimentally in charged colloidal
suspensions. Therefore, with a similar system as the one used by Imhof and
Dhont \cite{imhof95a,imhof95b,imhof97} the predictions of our model could
be rationalized experimentally.  On the other hand, our model considers
essentially the motion of mobile particles in a periodic potential. It
should be much simpler to develop analytic theories for this case than
for the transport of particles in a disordered potential energy landscape.

\section{The model and simulation details}
Charged particles (colloids) are modelled by an effective screened Coulomb
(or Yukawa) potential,
\begin{eqnarray}
\mbox{$V_{\rm Y}^{\rm {\alpha}{\beta}}$}\left(\mbox{$r$}\right)
= \varepsilon_{\rm {\alpha}{\beta}}
\frac{\exp(-\kappa{\sigma_{\rm {\alpha}{\beta}}}
\left(r\!/\!{\sigma_{\rm {\alpha}{\beta}}}\!-1\!\right))}
{\left(r\!/\!{\sigma_{\rm {\alpha}{\beta}}}\right)}
\label{Yukawapot}
\end{eqnarray}
where $r$ is the distance between particles $i$ and $j$. The parameters
$\sigma_{\rm {\alpha}{\beta}}$ ($\alpha, \beta={\rm s, l}$) denote
the distance between two particles at contact, $\sigma_{\alpha
\beta}=R_{\alpha}+R_{\beta}$, with $R_{\alpha}$ the radius of
an $\alpha$-particle. The interaction range is characterized by the
screening parameter $\kappa$. In the limit $\kappa \to 0$, the Yukawa
potential corresponds to bare Coulomb interactions.

A shifted Lennard--Jones (WCA) potential is used to model the excluded
volume of the particles:
\begin{eqnarray}
\mbox{$V_{\rm LJ}^{\rm {\alpha}{\beta}}$}\!\left(\mbox{$r$}\right)
=\left\{ \begin{array}{cr}
4\varepsilon\!\left[ {\left(\frac{\sigma_{\rm
{\alpha}{\beta}}}{r}\right)}^{\!12}- {\left(\frac{\sigma_{\rm
{\alpha}{\beta}}}{r}\right)}^{\!6}\,\right]
+ \varepsilon, & r\leq2^{\frac16} \sigma_{\rm {\alpha}{\beta}},
\label{shiftedLJ}\\ 0, & r > 2^{\frac16}\sigma_{\rm {\alpha}{\beta}}.
\end{array}\right.
\end{eqnarray}
The energy parameter $\varepsilon$ is set to $\varepsilon_{\rm ll}$.
Note that the potential, Eq.~(\ref{shiftedLJ}), is purely repulsive. It
can be regarded as an approximation to a hard--sphere interaction for
two particles at contact.

The binary Yukawa system of volume fraction $\phi=50\%$ consists of
$N_{\rm s}=820$ small and $N_{\rm l}=1640$ big particles in three
dimensional space. The size ratio between them is $\sigma_{\rm
ll}/\sigma_{\rm ss}\!=\!5$. We choose $\sigma_{\rm ll}=16.8$,
$\sigma_{\rm ss}=3.36$ and $\sigma_{\rm sl}=\left(R_{\rm s}+R_{\rm
l}\right)=10.08$, and mass $m_{\rm s}=4, m_{\rm l}=8$. The Yukawa
interactions are short--ranged with a fixed value of the inverse
screening parameter (screening length) $1/\kappa\!=\!0.119\sigma_{\rm
ll}=\!0.198\sigma_{\rm sl}=\!0.595\sigma_{\rm ss}$. Note that a
cut--off is introduced at $r_{\rm c}=3\sigma_{\rm ll}$ where potential
energies $V_{\rm Y}^{\rm {\alpha}{\beta}}\left(\mbox{$r_{\rm c}$}\right)$
are of the order of $10^{-7}$. For the energy parameters, the values
${\epsilon}_{\rm ll}=2.0$, ${\epsilon}_{\rm ss}={\epsilon}_{\rm ll}/625$,
and ${\epsilon}_{sl}={\epsilon}_{ll}/25$ are chosen. This choice implies
that small and large particles have the same surface charge density.
In the following,  all the physical quantities are measured in units of
the large particle's mass $m_{\rm l}=8$, diameter $\sigma_{\rm ll}$ and
energy ${\epsilon}_{\rm ll}$. The Boltzmann constant is set to $k_B=1.0$.

Newton's equations of motion for the charged particles are integrated
using the velocity form of the Verlet algorithm. The time step is chosen
to be $\delta t=0.015$ for $3.0{\geq}T{>}0.15$, $\delta t = 0.03$ for
$0.15{\geq}T{>}0.05$, and $\delta t = 0.045$ for $0.05{\geq}T{\geq}0.01$
(note that $\delta t$ is measured in units of the time $\tau_{\rm
MD}=\sigma_{\rm ll}\sqrt{\frac{m_{\rm l}}{{\epsilon}_{\rm ll}}}$).

At each temperature, the system was first equilibrated in the $NVT$
ensemble by coupling it to a stochastic heat bath. The equilibration
time was sufficiently longer than the time needed to observe a diffusive
motion of the small particles (typically, a factor of $4$ longer for
$T\ge0.15$, and a factor of $2$ for $0.03\le{T}<0.15$.) At the lowest
temperatures ($T<0.03$), the production runs were over 10 to 20 million
time steps. This was not enough to see a diffusive motion of the small
particles. Note that for the equilibration at a given temperature,
we used a final configuration of the next--higher temperature as an
initial configuration. Microcanonical production runs were performed at
the temperatures $T=3.0$, 2.5, 2.0, 1.0, 0.8, 0.5, 0.4, 0.2, 0.15, 0.1,
0.05, 0.04, 0.03, 0.025, 0.02, 0.015, 0.012, 0.01. At each temperature,
eight independent runs were done to improve statistics.

\section{Results}
\begin{figure}[t]
\twoimages[scale=0.3]{fig1abc.eps}{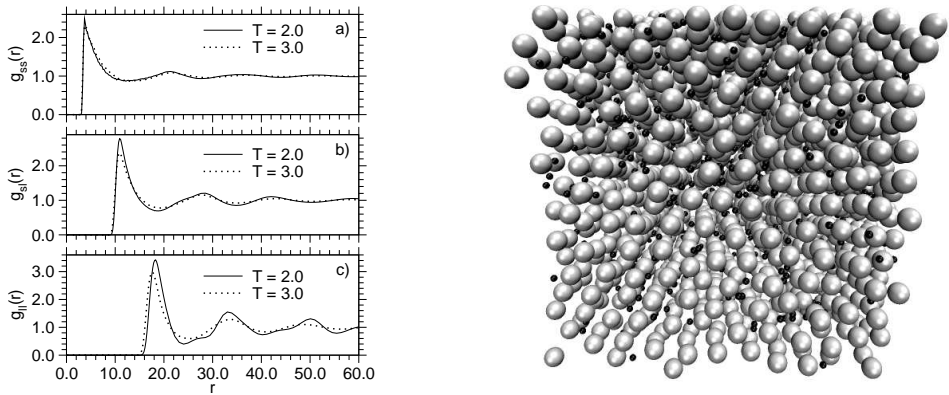}
\caption{Partial pair correlation function $g_{\alpha \beta}(r)$ for
the temperatures $T=3.0$ and $T=2.0$, i.e.~respectively above and below
the liquid--to--crystal transition of the large particles. a) $g_{\rm
ss}(r)$, b) $g_{\rm sl}(r)$, and c) $g_{\rm ll}(r)$. The snapshot shows
a configuration at $T=2.0$. Small and large particles are drawn as black
and gray spheres, respectively.}
\label{fig1}
\end{figure}
\begin{figure}[t]
\twoimages[scale=0.3]{fig2a.eps}{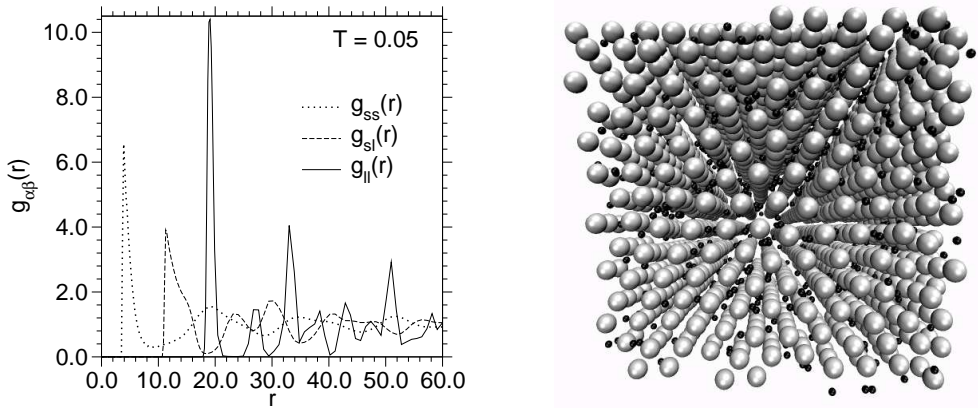}
\caption{Partial pair correlation function $g_{\alpha \beta}(r)$ for
temperature $T=0.05$. In the snapshot, small and large particles are
drawn as black and gray spheres, respectively.}
\label{fig2}
\end{figure}
First, we consider partial pair correlation functions \cite{hansen}
$g_{\alpha\beta}(r)$ ($\alpha, \beta = {\rm s, l}$) that measure
structural correlations between small (``s'' ) and large (``l'')
particles.  These quantities are proportional to the probability to
find a particle of type $\alpha$ at a distance $r$ from a particle of
type $\beta$.

\begin{figure}
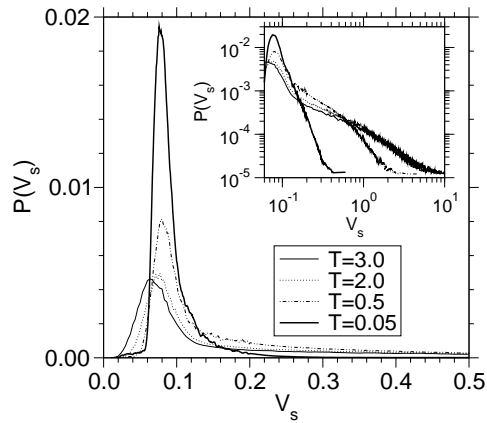

\onefigure[scale=0.3]{fig3.eps}
\caption{Distribution of single--particle potential energy of small
particles at different temperatures. The inset magnifies the high--energy
regime in a double--logarithmic plot.}
\label{fig3}
\end{figure}
\begin{figure}[t]
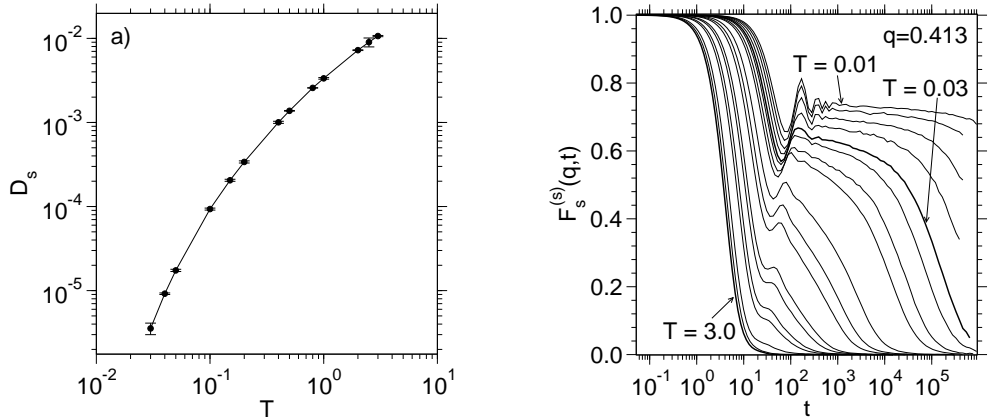

\twoimages[scale=0.30]{fig4a.eps}{fig4b.eps}
\caption{a) Self--diffusion constant of the small particles as a function
of temperature in double logarithmic plot.  b) Incoherent intermediate
scattering functions $F_{\rm s}(q,t)$ of the small particles at wavenumber
$q=0.413$ (corresponding to the first Bragg peak in the static structure
factor for the large-large correlations.)}
\label{fig4}
\end{figure}
\begin{figure}[t]
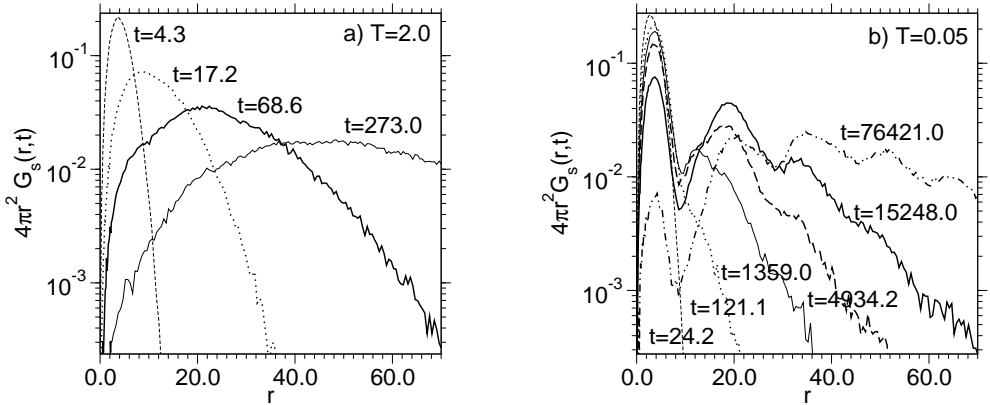

\twoimages[scale=0.30]{fig5a.eps}{fig5b.eps}
\caption{Self--part of the van Hove correlation function $G_{\rm s}(r,t)$
at various times for the temperatures a) $T=2.0$ and b) $T=0.05$.}
\label{fig5}
\end{figure}
Upon decreasing temperatures the large particles crystallize around
$T_{\rm m}\approx2.3$ into a fcc lattice structure. Fig.~\ref{fig1} shows
the $g_{\alpha \beta}(r)$ above and slightly below the freezing transition
of the large particles at $T=3.0$ and $T=2.0$, respectively. We see that
the curve for $g_{\rm ss}(r)$ at $T=3.0$ shows only tiny differences
to that at $T=2.0$. This also holds for the functions $g_{\rm sl}(r)$
at the two temperatures.  Beyond the first peak the amplitude in $g_{\rm
ss}(r)$ and $g_{\rm sl}(r)$ is only slightly different at $T=2.0$ and
at $T=3.0$.  This indicates a fluid--like behavior in the small--small
and small--large correlations. In contrast, the function $g_{\rm ll}(r)$
changes significantly from $T=2.0$ to $T=3.0$. Note that the rather broad
peaks at $T=2.0$ are reminiscent of a pair correlation function of a
normal liquid state. However, a closer inspection of other structural
quantities such as the static structure factor shows that the large
particles form a fcc crystal and the broad peaks in $g_{\rm ll}(r)$
are due to thermal fluctuations at the relatively high temperature
$T=2.0$. This can be also infered from the snapshot at $T=2.0$ (see
Fig.~\ref{fig1}). One can clearly identify the crystalline planes,
which are strongly distorted by the thermal motion of the large particles.

By decreasing temperature from $T=2.0$ to $T=0.05$, significant structural
changes occur. In Fig.~\ref{fig2}, both $g_{\rm ll}(r)$ and the snapshot
clearly show the long--range structural order of the fcc crystal at
$T=0.05$. The crystalline structure of the large particles is also
reflected in the correlations among the small particles at $T=0.05$. The
function $g_{\rm ss}(r)$ displays peaks at $r_{\rm ss}^{\rm (1)}\simeq3.9$
and $r_{\rm ss}^{\rm (2)}\simeq19.0$. The second peak at $r_{\rm ss}^{\rm
(2)}$ is at the same position as the first peak in $g_{\rm ll}(r)$, which
corresponds to the nearest neighbor distance in the fcc crystal.  This
indicates the tendency of the small particles to populate a sublattice
with respect to the fcc structure of the large particles. An interesting
feature is the rather sharp peak of $g_{\rm ss}(r)$ at $r_{\rm ss}^{\rm
(1)}$. This is due to the coordination of a small particle with another
small particle. It means that some of the small particles form pairs in
a given local site. Note that approximately 75\% of the small particles
have no nearest neighbor, 20\% exhibit a pairing, and 5\% of the small
particles are two--fold or three--fold coordinated. The occurrence of
the small particles with zero neighbour or pairs in local sites is also
reflected in $g_{\rm sl}(r)$. The first peak in this function shows
a shoulder around $r\simeq14.0$, which is approximately the distance
between a large particle and the next--nearest small particle of a small
particle pair.

The periodic structure of the large particles' crystalline matrix
is accompanied by fluid--like small particles, at least at high
temperatures. To characterize in detail the ``disorder'' in the
small--small correlations, we show in Fig.~\ref{fig3} the small particles'
distribution of single--particle potential energies (denoted by $P(V_{\rm
s})$) at four different temperatures. It is remarkable that the main peak
in $P(V_{\rm s})$ moves to higher energies while temperature decreases.
This means that the formation of the fcc structure by the large particles
is not energetically favoured by the small particles as it increases their
potential energy on average. Another significant feature of $P(V_{\rm
s})$ is the emergence of a high--energy tail. This is magnified in the
inset of Fig.~\ref{fig3} by a double--logarithmic plot of the data. This
high--energy tail tends to dissappear towards low temperatures. This
is due to the fact that the kinetic energy sets an upper bound for the
accessible potential energies.

Having discussed static properties of the model, we now turn our
attention to transport properties of the small particles in the
immobile environment of the large particles. Fig.~\ref{fig4}a displays
the small--particle self-diffusion constant $D_{\rm s}$ as a function
of temperature in a double--logarithmic plot (note that we determined
$D_{\rm s}$ from the mean--squared displacement of the small particles
via the Einstein relation \cite{hansen}). As temperature is decreased
from $T=3.0$ (i.e.~above the freezing transition of the large particles)
to $T=0.05$, the self-diffusion constant decreases by three and a half
orders of magnitude. Note that the functional form of $D_{\rm s}$ can
neither be described by a power law nor by an Arrhenius behavior.

A detailed description of the small--particle dynamics can be obtained
from the incoherent intermediate scattering functions $F_{\rm s}^{\rm
(s)}(q,t)$, the self--part of the time--dependent density--density
correlation function for the small particles \cite{hansen}. In
Fig.~\ref{fig4}b this function is shown for different temperatures
at $q=0.413$. This wavenumber corresponds to the location of the
first Bragg peak in the static structure factor for the large--large
correlations. Note that for other wavenumbers a similar behavior of
$F_{\rm s}^{\rm (s)}(q,t)$ is observed.  Above the liquid--to--crystal
transition of the large particles at $T_{\rm m}\approx2.3$, $F_{\rm
s}^{\rm (s)}(q,t)$ exhibits an exponential one--step decay to zero,
as expected for simple liquids. Below $T_{\rm m}$, a plateau develops
at intermediate times, followed by a {\it non--exponential} decay to
zero. The emergence of the plateau is associated with a caging of the
particles. In this case, the small particles are trapped in local
potential basins created by the large particles. The height of the
plateaus in $F_{\rm s}^{\rm (s)}(q,t)$ (also called Lamb--M\"ossbauer
factor \cite{glassbook}) measures the localization of the particles with
respect to the length scale corresponding to the considered wavenumber
$q$ \cite{glassbook}. We can infer from Fig.~\ref{fig4}b that around
$T_{\rm m}$ the small particles undergo a transition from a completely
delocalized state (with vanishing plateau) to a more and more localized
state towards low temperatures.  In the light of the potential energy
distribution (see Fig.~\ref{fig3}), the delocalization--to--localization
transition of the small particles is related to the gradual disappearance
of the high--energy tails.

The plateau sets in by an oscillation. With decreasing temperature,
this oscillation and the whole microscopic regime, i.e.~the decay of
the curves onto the plateau, shift to the right on the time axis. This
behavior corresponds to a softening of the cage towards lower temperatures
and is the subject of a forthcoming publication.

Further information on the long--time regime and the diffusive motion
of the small particles can be extracted from the Fourier transform of
$F_{\rm s}^{\rm (s)}(q,t)$, the self-part of the van Hove correlation
function, $G_{\rm s}^{\rm (s)}(r,t)$ \cite{hansen}.  $4\pi r^2G_{\rm
s}^{\rm (s)}(r,t)$ is the probability to find a particle at time $t$
at a distance $r$ away from the origin at $t=0$.  In Fig.~\ref{fig5},
this quantity is shown for different times at $T=2.0$ and $T=0.05$
in a linear--logarithmic plot. At the high temperature, we observe
a regular behavior, which is similar to that of simple liquids. With
increasing time, the location of the peaks moves continuously to larger
distances. However, the behavior is totally different at low temperatures:
At $t=24.2$, the correlation function exhibits a single peak. As time
goes on, a second peak develops and if one waits until $t=76421.0$ five
peaks can be identified. The distance between the peak maxima is around
$19.0$, which corresponds to the lattice constant of the large particle's
fcc lattice. We can conclude from this that at low temperatures the small
particles do not diffuse in a continuous manner but discontinuously in
time by hopping along the sublattice sites of the large particle's fcc
lattice. Note that this interpretation has been first given for similar
features in the glassy behavior of a soft--sphere model by Roux {\it
et al.}~\cite{roux89}

\section{Conclusions}
We have presented an extensive MD simulation study of a binary
Yukawa mixture with size ratio 1:5. In this system, interesting phase
behavior is observed where the crystallization of the large particles is
accompanied by a dynamic delocalization--to--localization transition of
the small particles.  We speculate that this kind of phase behavior is
generic for broad classes of complex systems, in particular mixtures of
disparate--sized (charged) colloids (see, e.g., the experiments by Imhof
and Dhont \cite{imhof95a,imhof95b,imhof97}). The motion of proteins in
lipid cubic phases \cite{tanaka04} might be also very similar to that
of the small particles in our Yukawa mixture.  However, we are not aware
of any systematic experiments on fluid--like particles in a crystalline
matrix.  We hope the present work will stimulate new experimental efforts
in this direction.

\acknowledgments
Financial support of the DFG (SFB 625, SFB TR6, and the Emmy Noether
programme, grant No. HO 2231/2) and the MWFZ Mainz are gratefully
acknowledged. We thank the NIC J\"ulich for a generous grant of computing
time on the JUMP.

\end{document}